\documentclass[10pt,twoside,BCOR7mm,DIV15,headinclude,footexclude,
               cleardoubleempty,idxtotoc]
{scrartcl}

\usepackage{natbib}
\usepackage[font=small,labelfont=bf]{caption}
\usepackage[english]{babel}
\usepackage{graphicx}
\usepackage{hyperref}
\usepackage{scrpage2}
\usepackage{ifthen}
\usepackage{booktabs}
\usepackage{amsmath}
\usepackage{amssymb}
\usepackage{multicol}
\usepackage{float}
\usepackage{hyperref}

\hypersetup{breaklinks=true
,colorlinks=true,linkcolor=black,urlcolor=blue
,citecolor=black}

\addto\captionsenglish{%
}

\makeatletter
\renewcommand{\@biblabel}[1]{}
\renewcommand{\@cite}[2]{%
{#1\ifthenelse{\boolean{@tempswa}}{,#2}{}}}
\makeatother
\ofoot{\thepage}
\ifoot{}

\setcapindent{0em}
\setheadsepline{1pt}

\setkomafont{pagehead}{\normalfont\sffamily}
\setkomafont{pagenumber}{\normalfont\rmfamily}

\makeatletter
\newcommand{\listofcontributions}{\@starttoc{con}}

\newcommand{\l@contribution} {\@dottedtocline{1}{1.5em}{2.3em}}
\makeatother

\newenvironment{contribution}{
\setcounter{section}{0}
\setcounter{figure}{0}
\setcounter{table}{0}
}{
\newpage
\lehead{}
\rohead{}
}

\begin{document}

\setlength{\baselineskip}{2.5ex}

\begin{contribution}
% RUNNING AUTHOR: PUT AUTHOR NAMED HERE
\lehead{S.\ Ram\'irez Alegr\'ia, A.-N.\ Chen\'e, J.\ Borissova, et al.}

% RUNNING TITLE; SHORTEN THE TITLE IF NECESSARY
% IN CASE OF A ONE-PAGE CONTRIBUTION (POSTER),
% SQUEEZE AUTHORS AND TITLE IN THIS LINE (Author: Title ...)
\rohead{Proceedings of ``International Workshop on Wolf-Rayet Stars''}

\begin{center}
% FULL TITLE HEADING
{\LARGE \bf A not so massive cluster hosting a very massive star}\\
\medskip

% AUTHORS LIST
{\it\bf S.\ Ram\'irez Alegr\'ia$^{1,2}$, A.-N.\ Chen\'e$^{3}$, J.\ Borissova$^{1,2}$, R.\ Kurtev$^{1,2}$, 
C.\ Navarro$^{1,2}$, M.\ Kuhn$^{2}$, J.A.\ Carballo-Bello$^{1,2}$}\\

% AFFILIATIONS
{\it $^1$Millennium Institute of Astrophysics, Santiago, Chile}\\
{\it $^2$Instituto de Astronom\'ia, Valpara\'iso, Chile}
{\it $^3$Gemini Observatory, AURA, USA}

% ABSTRACT
\begin{abstract}
We present the first physical characterization of the young open cluster VVV\,CL041. We  
spectroscopically observed the cluster main-sequence stellar population and a very-massive star
candidate: WR62-2. CMFGEN modeling to our near-infrared spectra indicates that WR62-2 is 
a very luminous (10$^{6.4\pm0.2} L_{\odot}$) and massive ($\sim80 M_{\odot}$) star. 
\end{abstract}
\end{center}

% TEXT OF THE PAPER, TWO-COLUMN STYLE
\begin{multicols}{2}

%\section{Results and analysis}

The current census of Wolf-Rayet (WR) in the Milky Way is far from complete (for example, there are $\sim$640 WR 
 stars reported in the online ``Galactic Wolf Rayet Catalogue'', but we expect over 1900 in our Galaxy;
 \citealt{rosslowe15}). It is also under debate the WR birth-place. We expect their formation in 
 clustered environments, but it is unclear whether it happens in young clusters or associations. Trying to 
 solve this question, we started a search of WR in the cluster candidates catalogue from \citet{borissova11}, 
 based on images from the ESO Public Survey VVV \citep{minniti10}). 

 Using VVV $JHK_S$ photometry and $H$-, $K$-band spectroscopy we determine the cluster distance 
(via spectroscopic parallax; $d = 4.2\pm0.9$ kpc), radius ($r=0.75'$), age (fitting Geneve-MS and PMS 
isochrones, \citealt{ekstrom12,siess00}; age=$1-5$ Myr), and total stellar mass (by integration of Kroupa-IMF, 
\citealt{kroupa01}, fitted to the cluster mass function; $M_{CL}=(3.1\pm0.6)\cdot10^3 M_{\odot}$).  
 
 The near-IR spectra also revealed part of the cluster main sequence population. We observe 8 stars, and 
 we classify 6 of them as OB-type stars (between O4\,V and early B-type). Spectrum of star \#8, the brightest
 star in the decontaminated CMD, displays the Brackett series with strong and broad emission lines. The C\,IV 
 and N\,III lines are clearly detected. The Brackett series in emission indicates that hydrogen is still present. 
 Carbon lines are narrow and weak, confirming that it is not a WC. We assigned a spectral type WN8-9h to star 
 \#8 (hereafter, WR62-2). 
 
Using the code CMFGEN \citep{hillier98}, we estimate for WR62-2 the effective temperature ($T_{eff}=34000 K$),
which indicates a luminosity of (10$^{6.4\pm0.2} L_{\odot}$) and a initial mass of at least $80 M_{\odot}$, from the
star position in the HR-diagram. The cluster and WR62-2 masses are incompatible with the ($M_{ecl} - m_{max}$)
relation \citep{weidner10}. A binary merge is a probable mechanism to explain the presence of this very massive star
in VVV\,CL041 and a gas remnant surrounding WR62-2 should be expected in mid-IR. 
%The WISE W3 image shows WR62-2 surrounded by a compact and bright cocoon.

%-----------One-column figure -----------------------------------
% Note that only the [H] option is allowed for placing 1-column figures!
\begin{figure}[H]
\begin{center}
\includegraphics[width=2.5cm]{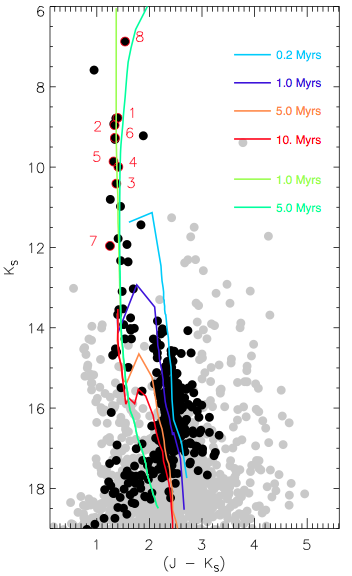}
\vspace{0.1cm}
\includegraphics[width=4.4cm]{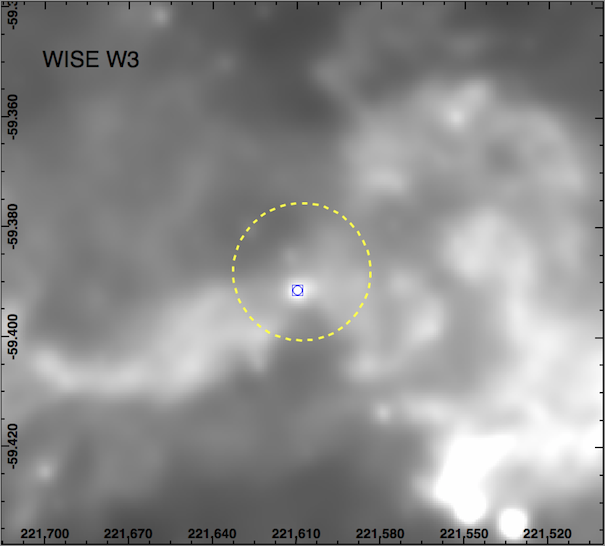}
\includegraphics[width=7.6cm]{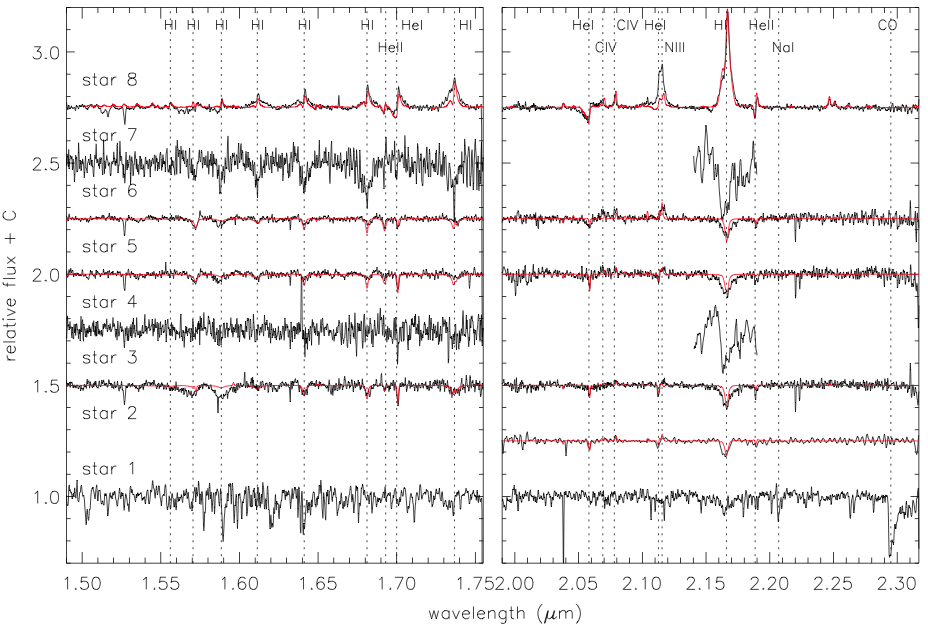}
\caption{{\it Top-left:} VVV\,CL041 field-star statistically decontaminated CMD. WR62-2 is shown on top of the cluster 
stellar population sequence. {\it Top-right:} VVV\,CL041 mid-IR WISE W3 image. The yellow circle and the blue square 
show the cluster's and WR62-2 positions. {\it Bottom:} Near-IR stellar spectra (WR62-2 is shown on top). The CMFGEN 
models are shown with red lines.
\label{example:smallfig}}
\end{center}
\end{figure}
%-----------------------------------------------------------

\bibliographystyle{aa} % style aa.bst
\bibliography{myarticle}

\end{multicols}

\end{contribution}

%%-------------------------------------------------------

\end{document}